\documentclass[a4paper,11pt]{article}

\usepackage{pos}

\title{
  The basics and applications of the tempered Lefschetz thimble method 
  for the numerical sign problem%
\footnote{Report No.: KUNS-2904}
}
\ShortTitle{Basics and applications of the tempered Lefschetz thimble method}

\author*[a]{Masafumi Fukuma}
\author[b]{Nobuyuki Matsumoto}

\affiliation[a]{
  Department of Physics, Kyoto University\\
  Kyoto 606-8502, Japan
}
\affiliation[b]{
  RIKEN/BNL Research center, Brookhaven National Laboratory,\\
  Upton, NY 11973, USA
}

\emailAdd{fukuma@gauge.scphys.kyoto-u.ac.jp}
\emailAdd{nobuyuki.matsumoto@riken.jp}

\abstract{
The numerical sign problem has long been a major obstacle 
to first-principles calculations 
in various important fields of physics. 
We report that the recently proposed algorithm, 
{\it tempered Lefschetz thimble method} (TLTM), 
and its worldvolume extension (WV-TLTM) 
can be a promising solution 
in its trustability and versatility.
}

\FullConference{%
 The 38th International Symposium on Lattice Field Theory, LATTICE2021
  26th-30th July, 2021
  Zoom/Gather@Massachusetts Institute of Technology
}

\newcommand{\bbC}{{\mathbb{C}}}
\newcommand{\bbR}{{\mathbb{R}}}

\newcommand{\ReS}{{\rm Re}\,S}
\newcommand{\ImS}{{\rm Im}\,S}

\begin{document}
\maketitle

\section{Introduction}
\label{sec:introduction}

The Monte Carlo (MC) method has been intensively used 
for first-principles calculations of the strong interaction, 
based on the lattice QCD. 
However, for the computation at finite density, 
the bosonized action becomes complex-valued, 
and one is forced to make a MC estimation 
of a highly oscillatory integral. 
This gives rise to an extraordinarily high computational cost, 
exponentially increasing with the degrees of freedom (DOF). 
Such a problem is called the {\it sign problem}, 
and has been a major obstacle 
to first-principles calculation 
in various important fields of physics. 
Typical examples are, 
in addition to the aforementioned finite density QCD \cite{Guenther:2021}, 
the $\theta$-vacuum with finite $\theta$, 
the Quantum Monte Carlo calculation 
of strongly correlated electron systems 
and frustrated spin systems, 
and the real-time dynamics of quantum fields. 
Among various approaches proposed so far 
towards solving the sign problem, 
in this talk we concentrate 
on the Lefschetz thimble method 
\cite{Cristoforetti:2012su,Fujii:2013sra,Alexandru:2015sua,
Fukuma:2017fjq,Alexandru:2017oyw,
Fukuma:2019wbv,Fukuma:2019uot,Fukuma:2020fez,Fukuma:2021aoo}, 
especially its tempered version \cite{Fukuma:2017fjq} 
as well as the worldvolume extension \cite{Fukuma:2020fez}. 

In section \ref{sec:LTM}, 
we explain the basics of the Lefschetz thimble method  
and point out that 
it suffers from the dilemma 
between the sign problem and the ergodicity problem. 
Section \ref{sec:TLTM_WV-TLTM} describes our solution, 
the {\it tempered Lefschetz thimble method} (TLTM) 
\cite{Fukuma:2017fjq}, 
and its improved version, 
the {\it worldvolume tempered Lefschetz thimble method} (WV-TLTM) 
\cite{Fukuma:2020fez}. 
The WV-TLTM is discussed in more detail 
in the contribution \cite{Fukuma:2021b} 
together with its statistical analysis method. 
In section \ref{sec:applications}, 
we exemplify the effectiveness of the (WV-)TLTM 
by its successful application to various models. 
Section \ref{sec:conclusion} is devoted to conclusion and outlook.%

\textbf{Note:} 
Some part of the presentation in this proceedings 
has a substantial overlap 
with a contribution to CCP2021 \cite{Fukuma:2021a}.

\section{Lefschetz thimble method}
\label{sec:LTM}

Our aim is to numerically estimate 
the expectation values of observables 
defined by the path integral 
\begin{align}
  \langle \mathcal{O}(x) \rangle 
  \equiv \frac{\int_{\bbR^N}dx\,e^{-S(x)}\,\mathcal{O}(x)}
  {\int_{\bbR^N}dx\,e^{-S(x)}},
\label{ev}
\end{align}
where $x=(x^i)\in\bbR^N$ is the dynamical variable, 
$S(x)$ the action, 
$dx\equiv\prod_{i=1}^N dx^i$ the measure of the path integral, 
and $\mathcal{O}(x)$ an observable of interest. 
When the action $S(x)$ is complex-valued, 
the Boltzmann weight 
$e^{-S(x)}/Z$ ($Z=\int dx\,e^{-S(x)}$) 
is not a positive semidefinite function, 
and thus cannot be regarded as a probability distribution. 
The simplest and most naive MC method for a complex action $S(x)$ 
is the {\it reweighting method}, 
where we use the real part $\ReS(x)$ for a new Boltzmann weight, 
and treat the phase factor $e^{-i\,\ImS(x)}$ as a part of observable: 
\begin{align}
  \langle \mathcal{O}(x) \rangle 
  = \frac{\langle e^{-i\,\ImS(x)}\,\mathcal{O}(x)\rangle_{\rm rewt}}
  {\langle e^{-i\,\ImS(x)}\rangle_{\rm rewt}}
  \quad 
  \Bigl(\langle f(x) \rangle_{\rm rewt}
  \equiv \frac{\int dx\,e^{-\ReS(x)}\,f(x)}
  {\int dx\,e^{-\ReS(x)}}\Bigr).
\label{rewt}
\end{align}
However, because of the highly oscillatory behavior of $e^{-i\,\ImS(x)}$ 
at large degrees of freedom (i.e., when $N\gg 1$),
Eq.~\eqref{rewt} becomes a ratio of very small quantities 
even when $\mathcal{O}(x)$ is an operator of $O(1)$: 
\begin{align}
  \langle \mathcal{O}(x) \rangle 
  = \frac{\langle e^{-i\,\ImS(x)}\,\mathcal{O}(x)\rangle_{\rm rewt}}
  {\langle e^{-i\,\ImS(x)}\rangle_{\rm rewt}}
  = \frac{e^{-O(N)}}{e^{-O(N)}}
  \,\bigl(\,= O(1)\bigr).
\label{ratio_rewt}
\end{align}
Of course, this does not cause a problem 
if we can evaluate both the numerator and the denominator precisely. 
However, in the MC calculations 
they are estimated separately from sample averages, 
and thus are necessarily accompanied by statistical errors, 
leading to an estimate of the form
\begin{align}
  \langle \mathcal{O}(x) \rangle 
  \approx
  \frac{e^{-O(N)} \pm O(1/\sqrt{N_{\rm conf}})}
  {e^{-O(N)} \pm O(1/\sqrt{N_{\rm conf}})}.
\end{align}
Thus, 
in order for the statistical errors to be smaller than the main parts, 
we need the inequality 
$\sqrt{N_{\rm conf}} \lesssim e^{-O(N)}$, 
which means that the sample size must be exponentially large: 
$
  N_{\rm conf} \gtrsim e^{O(N)}.
$
This is the sign problem.

\begin{figure}[t]
  \centering
  \includegraphics[width=50mm]{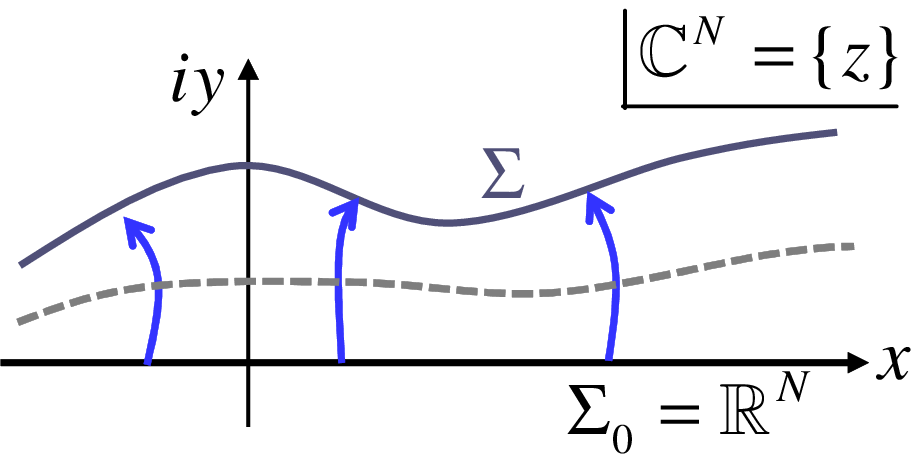}\hspace{10mm}
  \includegraphics[width=60mm]{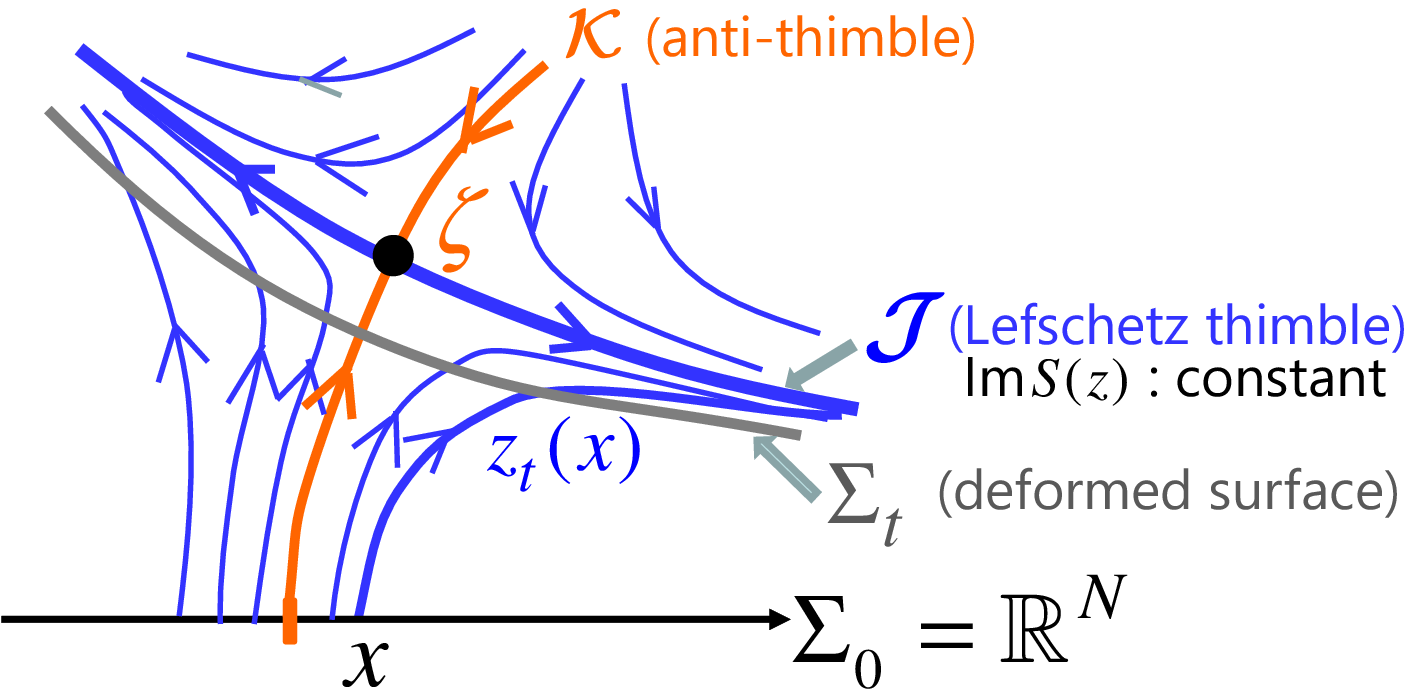}
  \caption{Deformation of integration surface.}
  \label{fig:LTM1-2}
\end{figure}

In the Lefschetz thimble method, 
we complexify the dynamical variable 
from $x=(x^i)\in\bbR^N$ to $z=(z^i)=x + i y\in\bbC^N$. 
We make an assumption 
(which usually holds for systems of interest) 
that $e^{-S(z)}$ and $e^{-S(z)}\,\mathcal{O}(z)$ 
are both entire functions over $\bbC^N$. 
Then, due to Cauchy's theorem, 
the integrals do not change under continuous deformations 
of integration surface 
(see the left panel of Fig.~\ref{fig:LTM1-2}):%
\begin{align}
  \langle\mathcal{O}(x)\rangle 
  = \frac{\int_{\bbR^N}dx\,e^{-S(x)}\,\mathcal{O}(x)}
  {\int_{\bbR^N}dx\,e^{-S(x)}}
  = \frac{\int_{\Sigma}dz\,e^{-S(z)}\,\mathcal{O}(z)}
  {\int_{\Sigma}dz\,e^{-S(z)}}. 
\label{cauchy}
\end{align}%
Thus, even when the sign problem is very severe on $\Sigma_0=\bbR^N$ 
due to the highly oscillatory phase factor $e^{-i\, \ImS(x)}$, 
the sign problem will be significantly reduced 
if the deformed integration surface $\Sigma$ can be chosen 
such that $\ImS(z)$ is almost constant on $\Sigma$.

The prescription for such a deformation 
is the following {\it anti-holomorphic gradient flow} 
(see the right panel of Fig.~\ref{fig:LTM1-2}):
\begin{align}
  \dot{z}_t = [\partial S(z_t)]^\ast,\quad
  z_{t=0} = x,
\label{flow_eq1}
\end{align}
which defines a map from $x\in\Sigma_0=\bbR^N$ 
to $z=z_t(x)\in\bbC^N$. 
We denote the deformed surface at flow time $t$ 
by $\Sigma_t\,(\equiv\{z_t(x)\,|\,x\in\bbR^N\})$. 
From the inequality 
\begin{align}
  [S(z_t)]^\cdot = \partial S(z_t)\cdot\dot{z}_t 
  = |\partial S(z_t)|^2 \geq 0,
\end{align}
we see that 
\begin{itemize}
\item[(a)] 
$[\ReS(z_t)]^\cdot\geq0$, 
i.e., $\ReS(z_t)$ always increases along the flow 
except at critical points 
(points where the gradient of the action vanishes), 
\vspace{-2mm}

\item[(b)]
$[\ImS(z_t)]^\cdot=0$, 
i.e., $\ImS(z_t)$ is constant along the flow. 
\end{itemize}

Associated with a critical point $\zeta$, 
we define the corresponding Lefschetz thimble $\mathcal{J}$ 
as a union of orbits flowing out of $\zeta$,%
\footnote{
  We here extend $z_t$ to be a map from $\bbC^N$ to $\bbC^N$ 
  by allowing an initial point $z_{t=0}$ to be in $\bbC^N$. 
} 
\begin{align}
  \mathcal{J}\equiv \bigl\{z\,\bigl|\,\lim_{t\to-\infty}z_t(z)=\zeta\bigr\}.
\end{align}
Due to the property (b), 
$\ImS(z)$ is constant on $\mathcal{J}$, 
$\ImS(z)=\ImS(\zeta)$ $(z\in\mathcal{J})$. 
Thus, when $\Sigma_t$ approaches a single Lefschetz thimble 
in the large flow time limit, 
we expect that 
the integration on $\Sigma_t$ becomes free from the sign problem 
by taking the flow time $t$ to be sufficiently large. 

The disappearance of the sign problem actually goes as follows. 
When we make a reweighting at flow time $t$, 
the main parts in the numerator and the denominator 
turn out to be of order $e^{-e^{-\lambda t}\,O(N)}$ 
with $\lambda$ a typical singular value of the matrix 
$\partial_i\partial_j S(\zeta)$, 
so that the expectation value is estimated as%
\footnote{ 
  $e^{i\theta(z)}\equiv e^{-i\,\ImS(z)}\,dz/|dz|$ 
  $(z\in\Sigma_t)$ 
  is the reweighting factor.
} 
\begin{align}
  \langle \mathcal{O}(x) \rangle 
  = \frac{\langle e^{i\theta(z)} \mathcal{O}(z)\rangle_{\Sigma_t}}
  {\langle e^{i\theta(z)}\rangle_{\Sigma_t}}
  \approx \frac{e^{-e^{-\lambda t}\,O(N)}\pm O(1/\sqrt{N_{\rm conf}})}
  {e^{-e^{-\lambda t}\,O(N)}\pm O(1/\sqrt{N_{\rm conf}})}.
\label{ltm}
\end{align}
Therefore, if we take the flow time $t$ 
to be sufficiently large so as to satisfy $e^{\lambda t}=O(N)$, 
the main parts become $O(1)$, 
and thus we no longer need a huge size of sample. 

\begin{figure}[t]
  \centering
  \includegraphics[width=55mm]{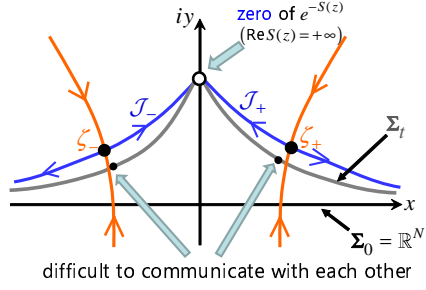}
  \caption{Ergodicity problem caused by a zero of $e^{-S(z)}$.}
  \label{fig:ergodicity_problem}
\end{figure}

So far, so good. 
However, when multiple thimbles are relevant to estimation, 
there generically arises another problem at large flow times, 
the ergodicity problem. 
Figure~\ref{fig:ergodicity_problem} is a sketch 
of the deformation of integration surface 
for the case $e^{-S(x)}=e^{-\beta x^2/2}\,(x-i)^\beta$. 
There, in addition to two critical points 
at $\zeta_\pm=\pm\sqrt{3}/2+i/2$ 
and the corresponding thimbles $\mathcal{J}_\pm$, 
we have a zero of $e^{-S(z)}$ at $z=i$, 
which behaves as an infinitely high potential barrier 
on $\Sigma_t$. 
Thus, in a Markov chain MC simulation, 
the configuration space $\Sigma_t$ is not well explored 
due to this potential barrier, 
and we eventually need a large computation time 
to realize global equilibrium. 
This is the ergodicity problem, 
which is believed to remain 
after taking the continuum limit \cite{Fujii:2015bua}. 

As a solution to this ergodicity problem, 
it was made a very interesting proposal in Ref.~\cite{Alexandru:2015sua} 
to employ a flow time 
which is sufficiently large so as to resolve the sign problem 
but at the same time 
is not too large so as to avoid the ergodicity problem. 
However, explicit simulations show that 
in many important cases the sign problem becomes mild 
only after the deformed surface reaches a zero, 
and furthermore, 
there is no reason to be capable of finding such a nice flow time 
for a system at large degrees of freedom, 
for which the flows around critical points and zeros are complicated.  

The {\it tempered Lefschetz thimble method} (TLTM) 
\cite{Fukuma:2017fjq} was proposed 
to resolve both
the sign problem (severe at small flow times) 
and the ergodicity problem (severe at large flow times) 
simultaneously 
by implementing the tempering algorithm 
\cite{Marinari:1992qd,Swendsen1986,Geyer1991,Hukushima1996}
to the Lefschetz thimble method. 

\section{Tempered Lefschetz thimble method and its worldvolume extension}
\label{sec:TLTM_WV-TLTM}

\subsection{Tempered Lefschetz thimble method}
\label{sec:TLTM}

The basic algorithm of the TLTM is as follows \cite{Fukuma:2017fjq}
(see the left panel of Fig.~\ref{fig:tltm}): 
\begin{figure}[t]
  \centering
  \includegraphics[width=55mm]{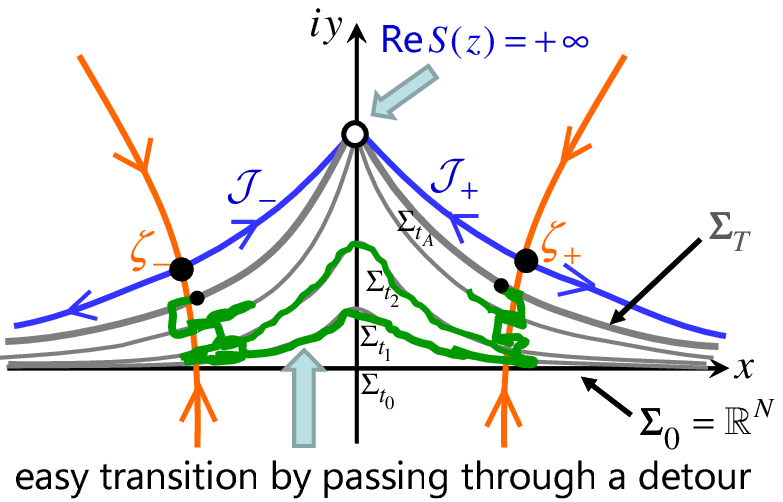} \hspace{5mm}
  \includegraphics[width=55mm]{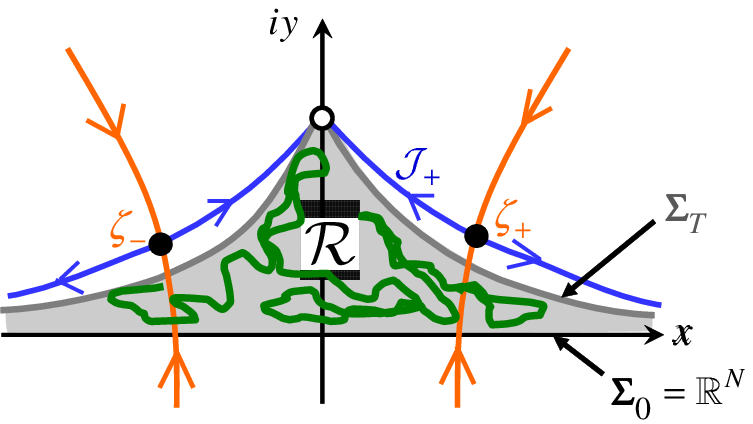}
  \caption{Configuration spaces of TLTM (left) and WV-TLTM (right).} 
  \label{fig:tltm}
\end{figure}%
\begin{itemize}
\item[(a)]
We set the target flow time $T$ 
which is large enough to resolve the sign problem. 

\item[(b)]
We introduce replicas in between 
the original integration surface $\Sigma_0=\bbR^N$ 
and the target integration surface $\Sigma_T$: 
$
  \{\Sigma_{t_0=0},\,\Sigma_{t_1},\,\Sigma_{t_2},\ldots,\Sigma_{t_A=T}\}.
$

\item[(c)]
We construct a Markov chain 
which consists of 
(i) transitions on each surface $\Sigma_{t_a}$ 
and (ii) exchanges of configurations between adjacent replicas 
$\Sigma_{t_a}$ and $\Sigma_{t_{a+1}}$. 
The Markov chain is designed 
so that the equilibrium distribution on the extended configuration space 
$\bbR^N\times\mathcal{A}=\{(x,t_a)\,|\,x\in\bbR^N,\,a=1,\ldots,A\}$ 
is given by 
$\rho(x,t_a)\propto e^{-S(z_{t_a}(x))}\,|\det J_{t_a}(x)|$, 
where $J_t(x)=\partial z_t(x)/\partial x$ 
is the Jacobian matrix of the flow. 

\item[(d)]
After equilibration, 
we estimate observables with a subsample from the replica at $t_A=T$. 
\end{itemize}

\noindent
Note that 
estimates from replicas at large flow times 
(where the sign problem is reduced) 
must agree within statistical errors due to Cauchy's theorem. 
This observation enables us to enhance the precision of estimate 
by using the $\chi^2$ fit 
with a constant as the fitting function 
\cite{Fukuma:2019wbv}. 

\subsection{Worldvolume Tempered Lefschetz Thimble Method (WV-TLTM)}
\label{sec:WV-TLTM}

Recall that in the original TLTM \cite{Fukuma:2017fjq}, 
we temper a system with the flow time 
introducing a finite discrete set of replicas 
$\{\Sigma_{t_0},\Sigma_{t_1},\ldots,\Sigma_{t_A}\}$.  
The advantage of this method over others is its versatility. 
In fact, the method can in principle be applied to any systems 
once the problem is defined 
in a path integral form over continuous variables. 
A drawback is its high computational cost at large DOF. 
In fact, 
the algorithm requires the computation of $\det J_t(x)$ 
in generating a configuration, 
whose cost is $O(N^3)$. 
This cost is further multiplied by the additional cost 
due to the implementation of the tempering, 
which is expected to be $O(N^{0-1})$. 

To overcome this drawback, 
the worldvolume tempered Lefschetz thimble method (WV-TLTM) was invented 
in Ref.~\cite{Fukuma:2020fez}, 
where HMC updates are performed 
on a continuous accumulation of integration surfaces, 
$\mathcal{R}\equiv \bigcup_{0\leq t\leq T}\Sigma_t$ 
(see the right panel of Fig.~\ref{fig:tltm}). 
We call this region $\mathcal{R}$ 
the {\it worldvolume} of integration surface.%
\footnote{ 
  We here borrow the terminology of string theory, 
  where, as an orbit of a particle is called a worldline, 
  that of a string is called a worldsheet, 
  and that of a membrane (that of a surface) a worldvolume.
} 
This WV-TLTM significantly reduces the numerical cost at large DOF, 
keeping the aforementioned advantages intact. 
In fact, in the WV-TLTM,
\begin{itemize}
\item
we no longer need to worry about the acceptance rate 
in the exchange process, 
\vspace{-3mm}

\item
we no longer need to compute the Jacobian of the flow 
in generating a configuration,
\vspace{-3mm}

\item
we can move configurations largely due to the use of the HMC algorithm.
\end{itemize}

The fundamental idea behind the WV-TLTM is 
that Cauchy's theorem allows us to average 
the denominator and the numerator in Eq.~\eqref{cauchy} 
over $t$ with an arbitrary weight $e^{-W(t)}$: 
\begin{align}
  \langle \mathcal{O}(x) \rangle
  &= \frac%
  {\int_{\Sigma_0}dx\,e^{-S(x)}\,\mathcal{O}(x)}
  {\int_{\Sigma_0}dx\,e^{-S(x)}}
  = \frac%
  {\int_{\Sigma_t}dz_t\,e^{-S(z_t)}\,\mathcal{O}(z_t)}
  {\int_{\Sigma_t}dz_t\,e^{-S(z_t)}}
\nonumber
\\
  &= \frac%
  {\int_0^T dt\,e^{-W(t)}\,\int_{\Sigma_t}dz_t\,e^{-S(z_t)}\,\mathcal{O}(z_t)}
  {\int_0^T dt\,e^{-W(t)}\,\int_{\Sigma_t}dz_t\,e^{-S(z_t)}}
  = \frac%
  {\int_{\mathcal{R}}dt\, dz_t\,e^{-W(t)-S(z_t)}\,\mathcal{O}(z_t)}
  {\int_{\mathcal{R}}dt\, dz_t\,e^{-W(t)-S(z_t)}}. 
\end{align}
The final expression does take a form of path integral 
over the region $\mathcal{R}$, 
and further can be rewritten to a ratio of reweighted integrals 
with a potential $V(z) \equiv \ReS(z)+W(t(z))$ 
[$t(z)$ is the flow time at $z\in\mathcal{R}$]. 
As can be seen from the absence of the Jacobian in $V(z)$, 
the molecular dynamics can be performed 
without calculating the Jacobian matrix \cite{Fukuma:2020fez} 
(see also the contribution \cite{Fukuma:2021b}). 
Although $e^{-W(t)}$ can be an arbitrary function of $t$ in principle, 
practically it is chosen 
such that the appearance ratios at different flow times $t$ 
are almost the same 
so as to ensure the region to be fully explored. 

\section{Applications}
\label{sec:applications}

The TLTM has been successfully applied to various models, 
including 
\begin{itemize}
\item
$(0+1)$-dimensional massive Thirring model \cite{Fukuma:2017fjq}
\vspace{-2mm}

\item
two-dimensional Hubbard model \cite{Fukuma:2019wbv,Fukuma:2019uot}
\vspace{-2mm}

\item
a chiral random matrix model (Stephanov model) \cite{Fukuma:2020fez}
\vspace{-2mm}

\item
antiferromagnetic Ising model on a triangular lattice 
[M.~Fukuma and N.~Matsumoto, talk at JPS meeting 2020]
\end{itemize}
Below we discuss the application of WV-TLTM \cite{Fukuma:2020fez} 
to the Stephanov model \cite{Stephanov:1996ki}.

\begin{figure}[t]
  \centering
  \includegraphics[width=120mm]{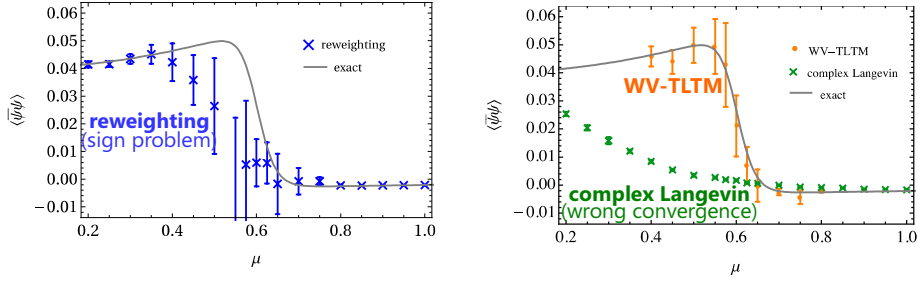}
  \caption{Chiral condensate $\langle\bar\psi\psi\rangle
  \equiv (1/2n)(\partial/\partial m)\ln Z_{\rm Steph}$ 
  \cite{Fukuma:2020fez}.}
  \label{fig:chiral_condensate}
\end{figure}

\begin{figure}[t]
  \centering
  \includegraphics[width=120mm]{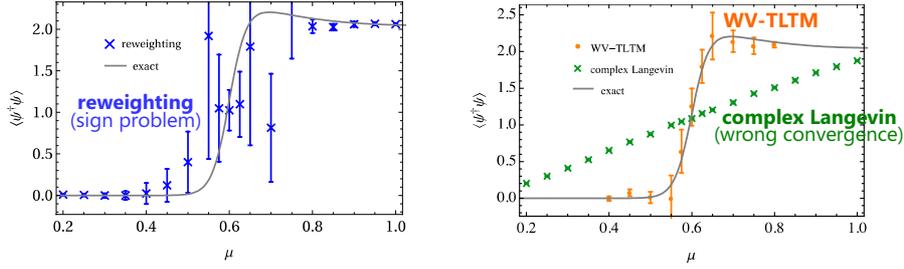}
  \caption{Baryon number density $\langle\psi^\dag\psi\rangle
  \equiv (1/2n)(\partial/\partial \mu)\ln Z_{\rm Steph}$ 
  \cite{Fukuma:2020fez}.}
  \label{fig:number_density}
\end{figure}

The grand partition function of finite density QCD takes the form 
\begin{align}
  Z_{\rm QCD} 
  =\int [dA_\mu]\,
  e^{(1/2g^2)\int {\rm tr}\,F_{\mu\nu}^2}\,
  {\rm Det}\,
  \left(\begin{array}{cc}
    m & \sigma_\mu(\partial_\mu+A_\mu)+\mu\\
    \sigma_\mu^\dag(\partial_\mu+A_\mu)+\mu & m
  \end{array}\right).
\end{align}
The Stephanov model is obtained 
by replacing the gauge-field degrees of freedom with a complex matrix, 
and takes the following form at zero temperature: 
\begin{align}
  Z_{\rm Steph} \equiv \int d^2 W\,e^{-n\, {\rm tr}\,W^\dag W}\,
  {\rm det}\,
  \left(\begin{array}{cc}
    m & i W+\mu\\
    i W^\dag+\mu & m
  \end{array}\right).
\end{align}
Here, $W=(W_{ij}=X_{ij}+i Y_{ij})$ is an $n\times n$ complex matrix, 
and thus the DOF is given by $N=2 n^2$. 
This corresponds to the DOF of the gauge field on a lattice, 
$4L^4(N_c^2-1)$ 
[$L$\,:\,\,linear size of four-dimensional periodic lattice, 
and $N_c$\,:\,\,color ($N_c=3$ for QCD) 
]. 
This model is regarded as an important benchmark 
for algorithms towards solving the sign problem, 
because
\begin{itemize}
\item
the model well describes the qualitative behavior of finite density QCD 
in the large $n$ limit,
\vspace{-2mm}

\item
the complex Langevin method \cite{Parisi:1983cs,Klauder:1983sp}
fails for this model 
because of the serious wrong convergence problem 
\cite{Bloch:2017sex}.

\end{itemize}
Figures~\ref{fig:chiral_condensate} and \ref{fig:number_density} 
show the estimates of the chiral condensate 
and the baryon number density 
obtained with three methods: 
(i) the naive reweighting method, 
(ii) the complex Langevin method, 
and (iii) the WV-TLTM \cite{Fukuma:2020fez}. 
We see that the naive reweighting method fails 
due to the sign problem, 
and also that the complex Langevin method 
gives wrong results with small statistical errors. 
On the other hand, 
the WV-TLTM gives correct results agreeing with the exact values. 
In fact, 
this is the first (and only at the moment) successful example 
among the attempts using various methods 
to reproduce the correct results for the Stephanov model 
in parameter regions where the sign problem is severe.

\section{Conclusion and Outlook}
\label{sec:conclusion}

The sign problem has been an obstacle to 
first-principles calculations in various important fields of physics, 
and a versatile solution has long been awaited. 
In this talk, 
we report that the (WV-)TLTM can be 
one of the most powerful candidate as a solution. 

Our group including the present authors 
has already started the full-scale application of the WV-TLTM 
to systems at large DOF, 
porting the code so as to run on a supercomputer. 
In parallel with research in this direction, 
we believe that 
it is still important to continue developing the algorithm itself 
for further efficiency at large DOF. 

The most important in the future development 
will be the application to the real-time dynamics of quantum fields. 
If a MC method is at hand for time-dependent quantum systems, 
the first-principles calculation will become possible 
for nonequilibrium systems, 
such as the very early Universe and heavy ion collision experiments.

\acknowledgments
The authors thank Yusuke Namekawa, Issaku Kanamori and Yoshio Kikukawa 
for useful discussions.
This work was partially supported by JSPS KAKENHI 
(Grand Numbers JP20H01900, JP18J22698) 
and by SPIRITS 
(Supporting Program for Interaction-based Initiative Team Studies) 
of Kyoto University (PI: M.F.). 
N.M.\ is supported by the Special Postdoctoral Researchers Program 
of RIKEN.


\end{document}